\documentclass[12pt]{article}

\usepackage{amssymb}
\usepackage{amsfonts}
\usepackage{graphicx}

\newtheorem{theorem}{Theorem}

\begin{document}

\author{A.Yu.Khrennikov\footnote{International Center for Mathematical Modelling in Physics and Cognitive Sciences, Linnaeus University,
V\"axj\"o, SE-351 95, Sweden, Andrei.Khrennikov@lnu.se}, 
S.V.Kozyrev\footnote{Steklov Mathematical Institute, Gubkina str. 8, 119991, Moscow, Russia, kozyrev@mi.ras.ru}, 
K. Oleschko\footnote{Titular Researcher, Centro de Geociencias,
Universidad Nacional Aut\'onoma de M\'exico (UNAM),
Campus UNAM Juriquilla, Blvd. Juriquilla 3001,
Quer\'etaro, Qro., C.P. 76230, M\'exico,
olechko@servidor.unam.mx}, \\ 
A. G. Jaramillo\footnote{Director del Desarrollo T\'ecnico,
TEMPLE, M\'exico, armandogj@orionearth.mx}, 
M. de   Jes\'us Correa L\'opez\footnote{Caracterizaci\'on de Yacimientos, Coordinaciуn de Diseno de Proyectos,
Activo de Producci\'on Ku Maloob Zaap, 2 nivel, sala1, Ed. Kaxan, Av. Contadores,
Km 4.5 Carretera Carmen Puerto Real, Cd. Del Carmen, Camp., M\'exico,
maria.jesus.correa@pemex.com}
}

\title{Application of $p$-adic analysis to time series}

\maketitle

\begin{abstract}
Time series defined by a $p$-adic pseudo-differential equation is investigated using the expansion of the time
series over $p$-adic wavelets. Quadratic correlation function is computed. This correlation function shows a degree--like
behavior and is locally constant for some time periods. It is natural to apply this kind of models for the investigation of avalanche
processes and punctuated equilibrium as well as fractal-like analysis of time series generated by measurement of pressure
in oil wells.
\end{abstract}

\section{Introduction}

The present paper is devoted to application of $p$-adic analysis to investigation of time series. Time series is a (stochastic) process defined on natural numbers which is understood as a discretization of a process defined on real (positive) numbers, i.e. the discretization is defined by the injection $\mathbb{N}\to\mathbb{R}_{+}$. In the present paper we consider real and complex valued stochastic processes.

The alternative way to obtain time series from a process $f$ of a real argument is to use averaging, i.e. to consider the process $F(n)$, $n\in\mathbb{N}$ (we use the agreement that natural numbers contain zero) of the form
\begin{equation}\label{averaging}
F(n)=\int_{n}^{n+1}f(x)dx.
\end{equation}

In the present paper we propose the following approach to investigation of time series. We will consider complex valued stochastic processes of $p$-adic argument.

The field $\mathbb{Q}_p$ of $p$-adic numbers is a completion of the field of rational numbers with respect to the $p$-adic norm, $|x|_p=p^{-\gamma}$, $x=p^{\gamma}{m\over n}$, $x\ne 0$, $m$ and $n$ are not divisible by $p$. Recall that $p$-adic numbers are in one to one correspondence
with series
$$
x=\sum_{i=\gamma}^{\infty}x_ip^i,\qquad x_i=0,\dots,p-1.
$$

We will investigate the correspondence between processes with real and $p$-adic arguments using the Monna map which maps $p$-adic numbers onto positive half--line surjectively
$$
\eta:\mathbb{Q}_p \to \mathbb{R}_+,
$$
\begin{equation}\label{Monna}
\eta:\sum_{i=\gamma}^{\infty} x_i p^{i} \mapsto
\sum_{i=\gamma}^{\infty} x_i p^{-i-1},\quad x_i=0,\dots,p-1,\quad
\gamma \in \mathbb{Z}.
\end{equation}
The Monna map is 1-Lipshitz, i.e. $|\eta(x)-\eta(y)|\le |x-y|_p$,  and therefore transforms large $p$-adic distances to large real distances.
This map conserves the measure (i.e. maps $p$-adic balls to real intervals with the same measure, where we use the Haar measure on $\mathbb{Q}_p$ with the normalization in which the unit ball has a measure one).

Moreover this map gives a one to one correspondence between the set of natural numbers $\mathbb{N}$ and the group of residues $\mathbb{Q}_p/\mathbb{Z}_p$, where the group of residues is understood as the group of fractions
\begin{equation}\label{QpZp}
z=\sum_{i=\gamma}^{-1}z_ip^i,\qquad z_i=0,\dots,p-1,\quad \gamma\in\mathbb{Z}_{-}
\end{equation}
(i.e. elements of $\mathbb{Q}_p$) with addition modulo one.

We introduce the discretization of functions of $p$-adic argument using the homomorphism map $\mathbb{Q}_p\to\mathbb{Q}_p/\mathbb{Z}_p$. This discretization is defined as follows: for $f(x)$, $x\in \mathbb{Q}$ we define the discretization $F(z)$, $z\in \mathbb{Q}_p/\mathbb{Z}_p$
$$
F(z)=\int_{|x|_p\le 1}f(z+x)d\mu(x),
$$
where $z$ is understood as a fraction given by (\ref{QpZp}).

After the application of the Monna map this $p$-adic discretization procedure reduces to the discretization of functions of positive argument using the formula (\ref{averaging}).

In the present text we discuss the application of $p$-adic analysis to time series of stochastic processes of real and $p$-adic argument. We will consider processes defined by $p$-adic pseudo-differential equations as in \cite{BikulovVolovich}, \cite{Bikulov} and will use wavelets for spectral analysis of these equations.
(See, e.g., \cite{wavelets}--\cite{Z3}, for general theory of $p$-adic pseudo-differential equations, $p$-adic stochastics and wavelets.)
 Application of the Monna map makes time in the discussed stochastic processes a real parameter.

We discuss the obtained process as a model of sandpile avalanche
processes with possible applications to biology (evolution theory -- the model of  punctuated equilibrium) and in geophysics --
$p$-adic wavelet analysis for time series generated as the result of measurement of pressure in oil wells. Such time series
have the internal hierarchic structure corresponding to the processes of generation of cascades of pressure spikes, a kind of
sandpile avalanche
processes. Hence, one can approximate such processes by using methods of multi-fractal analysis. $p$-Adics is one of a few
exactly solvable fractal models. Hence, its application to analysis of the time series for pressure in oil wells
can be useful at least at the level of theoretical modeling. In this note we elaborated the method of representation of the real
time series in the $p$-adic domain. We also presented discretization algorithms for $p$-adic wavelet analysis including
theory of $p$-adic fractional pseudo-differential operators (Vladimirov operators). The latter was applied to find the
$p$-adic fractional derivatives for the real time series obtained on the basis of pressure measurement in oil wells,
the Projects N 168638  of the SENER-CONACYT-Hidrocarburos Research Program.

The exposition of the present paper is as follows.

In Section 2 we recall the definitions of the Haar wavelet basis and the corresponding wavelet spaces with the discrete arguments.

In Section 3 we discuss $p$-adic wavelets.

In Section 4 we discuss $p$-adic pseudodifferential operators and their discretization.

In Section 5 we consider the discretization of a fractional $p$-adic Brownian motion and compute the corresponding quadratic correlation function.

In Section 6 we generalize the results of Section 5 for the case of different time scales.

In Section 7 we give a conclusion and discussion of our results.

\section{Discrete wavelet transform}

Let us discuss the discrete wavelet transform of time series. We consider time series as a complex valued function in $l^2(\mathbb{Z}).$ For introduction to wavelets see \cite{Daubechies}.

The Haar wavelet basis in $l^2(\mathbb{Z})$ can be constructed as follows. These wavelets can be considered as a restriction to $\mathbb{Z}$ of Haar wavelets on $\mathbb{R}$ (with positive scales)  with the regularization $\psi(\cdot)\mapsto\psi(\cdot+0)$.

The Haar wavelet basis $\{\psi_{jn}\}$ in $l^2(\mathbb{Z})$ is defined as follows
\begin{equation}\label{discrete_haar}
\psi_{jn}(x)=2^{-{j\over 2}}\psi(2^{-j}x-n),\quad n\in\mathbb{Z},\quad j> 0,
\end{equation}
$$
\psi(x)=\chi_{[0,1/2)}(x)-\chi_{[1/2,1)}(x),
$$
where $\chi_{[a,b)}$ is a characteristic function of $[a,b)$, $x\in\mathbb{Z}$.

The wavelet spaces in $l^2(\mathbb{Z})$ have the following form. The space $V_j$, $j\ge 0$, is a linear span of translations of the scaling function $\chi_{[n2^{j},(n+1)2^{j})}$, $n\in\mathbb{Z}$. In particular  $V_j\supset V_k$ for $j<k$.

The space $W_j$, $j>0$, is a linear span of $\psi_{jn}$, $n\in\mathbb{Z}$.

One has
$$
V_{j-1}=V_j\oplus W_j.
$$
Moreover the space $V_{j-1}\subset l^2(\mathbb{Z})$ is the span of $W_j$ and the characteristic function $\chi_{[0,2^{j-1})}$.

\bigskip

The formula for orthogonal projection to $V_j$ in $l^2(\mathbb{Z})$:
$$
P_j:\, l^2(\mathbb{Z})\to V_j,
$$
\begin{equation}\label{projection}
P_j:\, f(x)\mapsto {1\over 2^j}\sum_{n\in\mathbb{Z}}\chi_{[n2^{j},(n+1)2^{j})}(x)\sum_{l:n2^{j}\le l< (n+1)2^{j}} f(l).
\end{equation}
Note that for a fixed $x$ only one $n$ contributes to the above sum.

The proof of the above formula is as follows. In the space $V_j$ there is the orthonormal basis $\{2^{-{j\over 2}}\chi_{[n2^{j},(n+1)2^{j})}\}$, $n\in\mathbb{Z}$. Application of the sum over projections to vectors from this basis to a function $f$ gives the above formula.

Application of the projection $P_j$ to $f\in l^2(\mathbb{Z})$ gives the low frequency part $P_j f$ of $f$, the high frequency part of $f$ will be given by  $(1-P_j) f$.

\section{$p$-Adic discrete wavelet transform}

The basis of $p$-adic wavelets in ${L}^2({\mathbb Q}_p)$ has the form \cite{wavelets}
(see also \cite{W1}--\cite{WA} for details):
$$
\psi_{k;\,j n}(x)=p^{-j/2}\chi\big(p^{-1}k(p^{j}x-n)\big)
\Omega\big(|p^{j}x-n|_p\big), \quad x\in {\mathbb Q}_p.
$$
Here the index $k\in \{1,2,\dots,p-1\}$, $j\in {\mathbb Z}$, the index $n$ is an element of the quotient group ${\mathbb Q}_p/{\mathbb Z}_p$ understood as a rational number of the form
$$
n=\sum_{i=a}^{-1}n_ip^i,
$$
where $a\in\mathbb{Z}_{-}$ (negative integer), $n_i\in\{0,\dots,p-1\}$. The addition in ${\mathbb Q}_p/{\mathbb Z}_p$ can be understood as the addition modulo one of fractions of the above form.

The function $\chi$ is the additive character of the field ${\mathbb Q}_p$:
$$
\chi(x)=\exp\left(2\pi i \sum_{i=a}^{-1}x_ip^i\right),
$$
where $\sum_i x_ip^i$ contains the terms from the expansion of $x\in\mathbb{Q}_p$ over the degrees of $p$:
\begin{equation}\label{expansion}
x=\sum_{i=a}^{\infty}x_ip^i,\quad n_i=0,\dots,p-1.
\end{equation}
The function $\Omega(\cdot)$ is the characteristic function of $[0,1]\subset{\mathbb R}$ (therefore $\Omega(|\cdot|_p)$ is the characteristic function of $\mathbb{Z}_p$).

$p$-Adic spaces $V_j$ take the form of the spaces of functions of $p^{j}$--locally constant functions with compact support (i.e. spaces of compactly supported functions satisfying $f(\cdot)=f(\cdot+p^{j})$). The completion of the space $V_0$ can be identified with $l^2(\mathbb{Q}_p/\mathbb{Z}_p)$.

The projection in $L^{2}(\mathbb{Q}_p)$ to the completion of the space $V_j$ is the analog of discretization. This projection is given by the formula
\begin{equation}\label{pi_j}
(\Pi_jf)(x)=p^{-j}\int_{|y|_p\le p^{j}}f(x+y)d\mu(y)
\end{equation}
where $\mu$ is the Haar measure, $x\in \mathbb{Q}_p/p^{-j}\mathbb{Z}_p$ (i.e. $x$ can be considered as given by the terms in expansion (\ref{expansion}) with $i< -j$).

\medskip

Let us discuss the correspondence between the $p$-adic wavelets and the real Haar wavelets given by the Monna map.
For $p=2$ the Monna map maps the real Haar wavelets to the described above $p$-adic wavelets (for $p\ne 2$ we get the generalization of the Haar basis):
$$
\psi_{k;\,j n}(x)=\psi_{k;\,j \eta(n)}(\eta(x))
$$
where at the LHS of the above formula we have $p$-adic wavelets and at the RHS we have real wavelets. Let us note that for $p=2$ we have $k=1$ (and therefore we can omit this index).

\medskip

Application of the Monna map allows to consider time series (defined on natural numbers) as time series on the set of residues $\mathbb{Q}_p/\mathbb{Z}_p$.
In particular the basis (\ref{discrete_haar}) of Haar wavelets in $l^2(\mathbb{N})$ (i.e. the restriction of (\ref{discrete_haar}) to $n\ge 0$) becomes the basis of $p$-adic wavelets on $\mathbb{Q}_p/\mathbb{Z}_p$.

\medskip

The projection (\ref{projection}) (restricted to $l^2({\mathbb N})$ where in (\ref{projection})
we had $p=2$) after application of the Monna map $\eta$ takes the form
$$
P_j:\, f(x)\mapsto {1\over p^j}\sum_{l\in p^{-j}\mathbb{Z}_p/\mathbb{Z}_p} f\left(\eta(\eta^{-1}(x)+l)\right),
$$
$x\in \mathbb{N}$,  $p^{-j}\mathbb{Z}_p/\mathbb{Z}_p$ is the set
$$
\sum_{i=-j}^{-1} x_i p^{i},\quad x_i=0,\dots,p-1.
$$

\section{Discretization of $p$-adic pseudodifferential operators}

The Vladimirov operator of $p$-adic fractional differentiation (for $\alpha>0$) can be defined as
\begin{equation}\label{Dalpha}
D^{\alpha} f(x)={p^{\alpha}-1\over 1-p^{-1-\alpha}}
\int_{\mathbb{Q}_p}\frac{f(x)-f(y)}{|x-y|_p^{1+\alpha}}d\mu(y).
\end{equation}

The operator $D^{\alpha}$ in $L^{2}(\mathbb{Q}_p)$ does not change the diameter of local constancy, therefore one can define the action of this operator on $f\in l^2(\mathbb{Q}_p/\mathbb{Z}_p)$ (i.e. completion of $V_0$):
\begin{equation}\label{Dalpha1}
D^{\alpha} f(x)={p^{\alpha}-1\over 1-p^{-1-\alpha}}
\sum_{y\in \mathbb{Q}_p/\mathbb{Z}_p}\frac{f(x)-f(y)}{|x-y|_p^{1+\alpha}}.
\end{equation}
Here $x,y \in \mathbb{Q}_p/\mathbb{Z}_p$.

Using the Monna map $\eta$ one can define the action of the Vladimirov operator of fractional differentiation to $L^2(\mathbb{R}_+)$,
as follows:
\begin{equation}
\label{real}
\partial_p^{\alpha} f(x)=
\frac{p^{\alpha}-1}{1-p^{-1-\alpha}}
\int_{0}^{\infty}\frac{f(x)-f(y)}{|\eta^{-1}(x)-\eta^{-1}(y)|_p^{1+\alpha}}dy,
\end{equation}
where $\eta^{-1}$ is the inverse to $\eta$. The Haar wavelets are eigenvectors of $\partial_p^{\alpha}$.

Discretization of the above formula in analogue to (\ref{Dalpha1}) gives
\begin{equation}
\label{natural}
T_p^{\alpha} f(x)=
\frac{p^{\alpha}-1}{1-p^{-1-\alpha}}
\sum_{y=0}^{\infty}\frac{f(x)-f(y)}{|\eta^{-1}(x)-\eta^{-1}(y)|_p^{1+\alpha}}.
\end{equation}
Here $x\in\mathbb{N}$, $f\in l^2(\mathbb{N})$.

\begin{figure}[ptb]
\begin{center}
\label{FIGURE1}
\includegraphics[width=8cm]{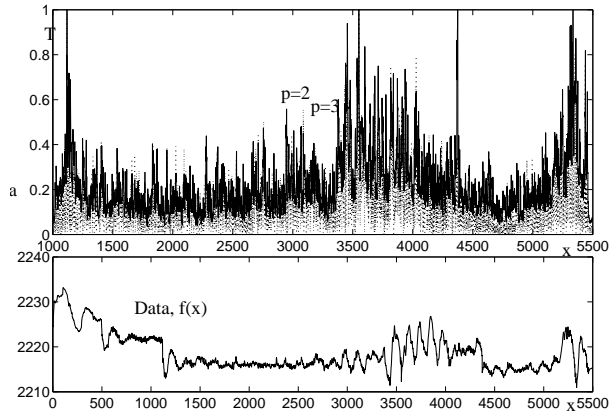}\caption{
$p$-Adic fractional derivatives ($p=2, 3$ and $\alpha=1)$ for the real time series obtained on the basis of pressure measurement in an oil well,
the Project N 168638 of the SENER-CONACYT-Hidrocarburos Research Program.}
\end{center}
\end{figure}

\section{Time series and $p$-adic Brownian motion}

$p$-Adic Brownian motion was introduced and the corresponding correlation functions were computed in \cite{BikulovVolovich}. Fractional $p$-adic Brownian motion was investigated in \cite{Bikulov}. Here we compute the quadratic correlation function for discretized fractional $p$-adic Brownian motion defined on $\mathbb{Q}_p/\mathbb{Z}_p$. We use the expansion of a stochastic process of $p$-adic argument over wavelets as in paper \cite{randomfield} (where actually a general locally compact ultrametric space was considered and the process under consideration was more general). The approach of the present paper differs from the mentioned above papers by the using of the discretization and application of expansion (\ref{wavelet_expansion}).

The fractional $p$-adic Brownian motion is a solution of the equation
\begin{equation}\label{brownian}
D^{\alpha}f(x)=\phi(x),
\end{equation}
where $\phi$ is the white noise (delta--correlated Gaussian mean zero generalized complex valued stochastic process over $\mathbb{Q}_p$). For discussion of $p$-adic generalized functions see \cite{Al-Kh-Sh=book}.

The white noise possesses the expansion over wavelets
$$
\phi(x)=\sum_{k;\,j n}d_{k;\,j n}\psi_{k;\,j n}(x)
$$
where $d_{k;\,j n}$ are mean zero Gaussian independent delta--correlated random variables.

The solution of (\ref{brownian}) in $D'(\mathbb{Q}_p)$ is given by the expansion over wavelets
\begin{equation}\label{wavelet_expansion}
f(x)=f_0+\sum_{k;\,j n}p^{-\alpha(1-j)}d_{k;\,j n}\left(\psi_{k;\,j n}(x)-\int_{\mathbb{Z}_p}\psi_{k;\,j n}(x)d\mu(x)\right)
\end{equation}
where $f_0$ is a mean zero Gaussian random variable independent from $d_{k;\,j n}$.

We will consider the solution with $f_0=0$, i.e. the solution which satisfies the initial condition
$$
\int_{\mathbb{Z}_p}f(x)d\mu(x)=0.
$$

Discretization $F=\Pi_0 f$ of $f$ given by (\ref{wavelet_expansion}) belongs to the space of linear functionals over $V_0$. We call this discretization the time series over
$\mathbb{Q}_p/\mathbb{Z}_p$.

\begin{theorem}\label{correlatorF}
The quadratic correlation function for the discretization $F=\Pi_0 f$ of (\ref{wavelet_expansion}) with the initial condition $F(0)=0$ is given by (\ref{correlator1}), where $\rho(0)=0$ and for $x\ne 0$ the function $\rho(x)$ is given by (\ref{rho1}).

\begin{equation}\label{correlator1}
\langle \overline{F(x)}F(y)\rangle=\rho(x)+\rho(y)-\rho(x-y),\qquad x,y\in \mathbb{Q}_p/\mathbb{Z}_p.
\end{equation}

\begin{equation}\label{rho1}
\rho(x)={1-p^{-1}\over 1-p^{2\alpha-1}}+|x|_p^{2\alpha-1}{p^{-2\alpha} -1\over 1-p^{2\alpha-1}}.
\end{equation}

\end{theorem}

\noindent{\bf Remark}\quad Application of the Monna map $\eta:\mathbb{Q}_p/\mathbb{Z}_p\to\mathbb{N}$ defines the equivalent stochastic process of natural argument, defined by the integral equation
\begin{equation}\label{TF=delta}
T_p^{\alpha} F(x)=\delta(x),\qquad x\in \mathbb{N},
\end{equation}
where the operator $T_p^{\alpha}$ is defined by (\ref{natural}) and $\delta(x)$ is the white noise on $\mathbb{N}$ (i.e. a set of delta--correlated Gaussian mean zero random variables).

\bigskip

\noindent{\it Proof}\qquad We perform the discretization of the expansion (\ref{wavelet_expansion}) by averaging over the ball $A$ of the diameter one. These balls are in one to one correspondence with the elements of $\mathbb{Q}_p/\mathbb{Z}_p$. Then we compute the correlation function for $F(A)$, $F(B)$
$$
F(A)=\int_{A}f(x)d\mu(x)
$$
where the averaging is taken with respect to the Haar measure $\mu$ over the balls $A$ and $B$.

$$
F(A)=\sum_{k;\,j n: A\le B_{jn}\le \,{\rm sup}\,(A,\mathbb{Z}_p)}p^{-\alpha(1-j)}d_{k;\,j n}\int_{A}\psi_{k;\,j n}(x)d\mu(x)-
$$
$$
-\mu(A)\sum_{k;\,j n: \mathbb{Z}_p\le B_{jn}\le \,{\rm sup}\,(A,\mathbb{Z}_p)}p^{-\alpha(1-j)}d_{k;\,j n}\int_{\mathbb{Z}_p}\psi_{k;\,j n}(x)d\mu(x)=
$$
$$
=\sum_{k;\,j n: A< B_{jn}< \,{\rm sup}\,(A,\mathbb{Z}_p)}p^{-\alpha(1-j)}d_{k;\,j n}\int_{A}\psi_{k;\,j n}(x)d\mu(x)-
$$
$$
-\mu(A)\sum_{k;\,j n: \mathbb{Z}_p< B_{jn}< \,{\rm sup}\,(A,\mathbb{Z}_p)}p^{-\alpha(1-j)}d_{k;\,j n}\int_{\mathbb{Z}_p}\psi_{k;\,j n}(x)d\mu(x)+
$$
$$
+\sum_{k;\,j n: B_{jn}= \,{\rm sup}\,(A,\mathbb{Z}_p)}p^{-\alpha(1-j)}d_{k;\,j n}\left(\int_{A}\psi_{k;\,j n}(x)d\mu(x)-\mu(A)\int_{\mathbb{Z}_p}\psi_{k;\,j n}(x)d\mu(x)\right).
$$
Here $B_{jn}$ is a ball where the wavelets $\psi_{k;\,j n}$, $k=1,\dots,p-1$ are supported.
The terms with $B_{jn}=A$ and $B_{jn}=\mathbb{Z}_p$ in the summation may be omitted.
In the following $\mu(A)=\mu(B)=1$ and $A,B\ne \mathbb{Z}_p$ (recall $F(\mathbb{Z}_p)=0$).

\bigskip

Consider the following cases.

A) $\langle \overline{F(A)}F(A)\rangle$, $|x|_p$, $x\in A$ is the distance between $A$ and zero.
$$
\langle \overline{F(A)}F(A)\rangle =\sum_{k;\,j n: A< B_{jn}< \,{\rm sup}\,(A,\mathbb{Z}_p)}p^{-2\alpha(1-j)}\left|\int_{A}\psi_{k;\,j n}(x)d\mu(x)\right|^2+
$$
$$
+\sum_{k;\,j n: \mathbb{Z}_p< B_{jn}< \,{\rm sup}\,(A,\mathbb{Z}_p)}p^{-2\alpha(1-j)}\left|\int_{\mathbb{Z}_p}\psi_{k;\,j n}(x)d\mu(x)\right|^2+
$$
$$
+\sum_{k;\,j n: B_{jn}= \,{\rm sup}\,(A,\mathbb{Z}_p)}p^{-2\alpha(1-j)}\left|\int_{A}\psi_{k;\,j n}(x)d\mu(x)-\int_{\mathbb{Z}_p}\psi_{k;\,j n}(x)d\mu(x)\right|^2=
$$
$$
=\sum_{1< p^{j}< |x|_p}p^{-2\alpha(1-j)}(p-1)p^{-j}+\sum_{1< p^{j}< |x|_p}p^{-2\alpha(1-j)}(p-1)p^{-j}+2p^{-2\alpha(1-j)}p^{1-j}\bigr|_{p^{j}=|x|_p}=
$$
$$
=2(1-p^{-1})\sum_{1< p^{j}< |x|_p}p^{(2\alpha-1)(j-1)}+2p^{(2\alpha-1)(j-1)}\bigr|_{p^{j}=|x|_p}=
$$
$$
=2(1-p^{-1})\sum_{1< p^{j}\le |x|_p}p^{(2\alpha-1)(j-1)}+2p^{-1}p^{(2\alpha-1)(j-1)}\bigr|_{p^{j}=|x|_p}=
$$
$$
=2(1-p^{-1})\sum_{1\le p^{j}< |x|_p}p^{(2\alpha-1)j}+2p^{-1}{|x|_p^{2\alpha-1}\over p^{2\alpha-1}}=2(1-p^{-1}){1-|x|_p^{2\alpha-1}\over 1-p^{2\alpha-1}}+2p^{-1}{|x|_p^{2\alpha-1}\over p^{2\alpha-1}}=
$$
$$
=  2\left({1-p^{-1}\over 1-p^{2\alpha-1}}+
|x|_p^{2\alpha-1}{p^{-2\alpha} -1\over 1-p^{2\alpha-1}}\right).
$$

Here we use the identity
$$
p^{-j}\sum_{k=1,\dots,p-1}\left|e^{2\pi i k/p^{-1}}-1\right|^2=p^{-j}\sum_{k=1,\dots,p-1}\left(2- e^{2\pi i k/p^{-1}}-e^{-2\pi i k/p^{-1}}\right)=
$$
$$
=p^{-j}\left(2(p-1)+1-\sum_{k=0,\dots,p-1}e^{2\pi i k/p^{-1}}+1-\sum_{k=0,\dots,p-1}e^{-2\pi i k/p^{-1}}\right)=2p^{1-j}.
$$

B) $\langle \overline{F(A)}F(B)\rangle$, the distance from $\mathbb{Z}_p$ to $A$ (denoted by $|x|_p$ for $x\in A$) is smaller than the distance $|y|_p$, $y\in B$ from $\mathbb{Z}_p$ to $B$.
$$
\langle \overline{F(A)}F(B)\rangle =
\sum_{k;\,j n: \mathbb{Z}_p< B_{jn}< \,{\rm sup}\,(A,\mathbb{Z}_p)}p^{-2\alpha(1-j)}\left|\int_{\mathbb{Z}_p}\psi_{k;\,j n}(x)d\mu(x)\right|^2-
$$
$$
-\sum_{k;\,j n: B_{jn}= \,{\rm sup}\,(A,\mathbb{Z}_p)}p^{-2\alpha(1-j)}\left(\overline{\int_{A}\psi_{k;\,j n}(x)d\mu(x)-\int_{\mathbb{Z}_p}\psi_{k;\,j n}(x)d\mu(x)}\right)\int_{\mathbb{Z}_p}\psi_{k;\,j n}(x)d\mu(x).
$$

Using the identity
$$
p^{-j}\sum_{k=1,\dots,p-1}\left(e^{-2\pi i k/p^{-1}}-1\right)=p^{-j}\left(1-p +\sum_{k=0,\dots,p-1}e^{-2\pi i k/p^{-1}}-1\right)=-p^{1-j}
$$
we get
$$
\langle \overline{F(A)}F(B)\rangle =\sum_{1< p^{j}< |x|_p}p^{-2\alpha(1-j)}(p-1)p^{-j}+p^{-2\alpha(1-j)}p^{1-j}\bigr|_{p^{j}=|x|_p}=
$$
$$
=(1-p^{-1})\sum_{1< p^{j}< |x|_p}p^{(2\alpha-1)(j-1)}+p^{(2\alpha-1)(j-1)}\bigr|_{p^{j}=|x|_p}=
$$
$$
=(1-p^{-1})\sum_{1< p^{j}\le |x|_p}p^{(2\alpha-1)(j-1)}+p^{-1}p^{(2\alpha-1)(j-1)}\bigr|_{p^{j}=|x|_p}=
$$
$$
=(1-p^{-1})\sum_{1\le p^{j}< |x|_p}p^{(2\alpha-1)j}+p^{-1}{|x|_p^{2\alpha-1}\over p^{2\alpha-1}}=(1-p^{-1}){1-|x|_p^{2\alpha-1}\over 1-p^{2\alpha-1}}+p^{-1}{|x|_p^{2\alpha-1}\over p^{2\alpha-1}}.
$$

C) $\langle \overline{F(A)}F(B)\rangle$, distance between  $A$ and $B$ (denoted by $|x-y|_p$, $x\in A$, $y\in B$) is smaller than $|x|_p$ (distance from $A$ to $\mathbb{Z}_p$).
$$
\langle \overline{F(A)}F(B)\rangle =\sum_{k;\,j n: \,{\rm sup}\,(A,B)< B_{jn}< \,{\rm sup}\,(A,\mathbb{Z}_p)}p^{-2\alpha(1-j)}\left|\int_{A}\psi_{k;\,j n}(x)d\mu(x)\right|^2+
$$
$$
+\sum_{k;\,j n: \mathbb{Z}_p< B_{jn}< \,{\rm sup}\,(A,\mathbb{Z}_p)}p^{-2\alpha(1-j)}\left|\int_{\mathbb{Z}_p}\psi_{k;\,j n}(x)d\mu(x)\right|^2+
$$
$$
+\sum_{k;\,j n: B_{jn}= \,{\rm sup}\,(A,\mathbb{Z}_p)}p^{-2\alpha(1-j)}\left|\int_{A}\psi_{k;\,j n}(x)d\mu(x)-\int_{\mathbb{Z}_p}\psi_{k;\,j n}(x)d\mu(x)\right|^2+
$$
$$
+\sum_{k;\,j n: B_{jn}= \,{\rm sup}\,(A,B)}p^{-2\alpha(1-j)}\overline{\int_{A}\psi_{k;\,j n}(x)d\mu(x)}\int_{B}\psi_{k;\,j n}(x)d\mu(x)=
$$
$$
=\sum_{|x-y|_p< p^{j}< |x|_p}p^{-2\alpha(1-j)}(p-1)p^{-j}+\sum_{1< p^{j}< |x|_p}p^{-2\alpha(1-j)}(p-1)p^{-j}+
$$
$$
+2p^{-2\alpha(1-j)}p^{1-j}\bigr|_{p^{j}=|x|_p}-p^{-2\alpha(1-j)}p^{-j}\bigr|_{p^{j}=|x-y|_p}=
$$
$$
=-(1-p^{-1})\sum_{1< p^{j}\le |x-y|_p}p^{(2\alpha-1)(j-1)}+2(1-p^{-1})\sum_{1< p^{j}< |x|_p}p^{(2\alpha-1)(j-1)}+
$$
$$
+2p^{(2\alpha-1)(j-1)}\bigr|_{p^{j}=|x|_p}-p^{-1}p^{(2\alpha-1)(j-1)}\bigr|_{p^{j}=|x-y|_p}=
$$
$$
=2(1-p^{-1})\sum_{1< p^{j}\le |x|_p}p^{(2\alpha-1)(j-1)}+2p^{-1}p^{(2\alpha-1)(j-1)}\bigr|_{p^{j}=|x|_p}-
$$
$$
-(1-p^{-1})\sum_{1< p^{j}\le |x-y|_p}p^{(2\alpha-1)(j-1)}-p^{-1}p^{(2\alpha-1)(j-1)}\bigr|_{p^{j}=|x-y|_p}=
$$
$$
=2(1-p^{-1})\sum_{1\le p^{j}< |x|_p}p^{(2\alpha-1)j}+2p^{-1}{|x|_p^{2\alpha-1}\over p^{2\alpha-1}}-(1-p^{-1})\sum_{1\le p^{j}< |x-y|_p}p^{(2\alpha-1)j}-p^{-1}{|x-y|_p^{2\alpha-1}\over p^{2\alpha-1}}=
$$
$$
=2(1-p^{-1}){1-|x|_p^{2\alpha-1}\over 1-p^{2\alpha-1}}+2p^{-1}{|x|_p^{2\alpha-1}\over p^{2\alpha-1}}-(1-p^{-1}){1-|x-y|_p^{2\alpha-1}\over 1-p^{2\alpha-1}}-p^{-1}{|x-y|_p^{2\alpha-1}\over p^{2\alpha-1}}.
$$

Let us denote (for $x\ne 0$, where $x\in\mathbb{Q}_p/\mathbb{Z}_p$)
\begin{equation}\label{rho}
\rho(x)=(1-p^{-1}){1-|x|_p^{2\alpha-1}\over 1-p^{2\alpha-1}}+p^{-1}{|x|_p^{2\alpha-1}\over p^{2\alpha-1}}={1-p^{-1}\over 1-p^{2\alpha-1}}+
|x|_p^{2\alpha-1}{p^{-2\alpha} -1\over 1-p^{2\alpha-1}}.
\end{equation}
Then the expression for the correlation function in the case (C) can be put in the form (recall that in this case $|x|_p=|y|_p$)
\begin{equation}\label{correlator}
\langle \overline{F(x)}F(y)\rangle=\rho(x)+\rho(y)-\rho(x-y),\qquad x,y\in \mathbb{Q}_p/\mathbb{Z}_p.
\end{equation}

In the case (B) the correlation function reduces to the same expression (where  $|y|_p=|x-y|_p$, therefore the corresponding terms in the correlator cancel). In the case (A) (when $x=y$ in $\mathbb{Q}_p/\mathbb{Z}_p$) we will obtain again the same expression for the correlator if we put $\rho(0)=0$.

Summing up, the quadratic correlation function for time series (the averaged Brownian motion) is given by (\ref{correlator}), where $\rho(0)=0$ and for $x\ne 0$ the $\rho(x)$ is given by (\ref{rho}).

This finishes the proof of the theorem. $\Box$

\section{Time series with different time scales}

In the present section we consider $p$-adic time series given by application of the operator $\Pi_l$ given by (\ref{pi_j}) to $p$-adic Brownian motion.

Let us consider the solution of (\ref{brownian}) given by the following modification of (\ref{wavelet_expansion}):
\begin{equation}\label{wavelet_expansion1}
f_{l}(x)=\sum_{k;\,j n}p^{-\alpha(1-j)}d_{k;\,j n}\left(\psi_{k;\,j n}(x)-p^{l}\int_{p^{l}\mathbb{Z}_p}\psi_{k;\,j n}(x)d\mu(x)\right).
\end{equation}

This solution satisfies the initial condition
$$
\int_{p^{l}\mathbb{Z}_p}f_l(x)d\mu(x)=0.
$$

The following theorem theorem is a generalization of Theorem \ref{correlatorF} and can be proven in analogous way.

\begin{theorem}\label{correlatorF_l}
The quadratic correlation function for the discretization $F_l=\Pi_l f$ of (\ref{wavelet_expansion1}) with the initial condition $F_l(0)=0$ is given by (\ref{correlator2}), where $\rho_l(0)=0$ and for $x\ne 0$ the function $\rho_l(x)$ is given by (\ref{rho2}).

\begin{equation}\label{correlator2}
\langle \overline{F_l(x)}F_l(y)\rangle=\rho_l(x)+\rho_l(y)-\rho_l(x-y),\qquad x,y\in \mathbb{Q}_p/p^{l}\mathbb{Z}_p.
\end{equation}

\begin{equation}\label{rho2}
\rho_l(x)=p^{-l}{1-p^{-1}\over 1-p^{2\alpha-1}}+|x|_p^{2\alpha-1}{p^{-2\alpha} -1\over 1-p^{2\alpha-1}}.
\end{equation}

\end{theorem}

In the limit $l\to\infty$ the first term (which does not depend on $x$) in (\ref{rho2}) vanishes.

\section{Concluding remarks and future challenges}

By Theorem \ref{correlatorF} and the Remark after this theorem we get time series on $\mathbb{N}$ defined by stochastic equation (\ref{TF=delta}).
Correlation function of this time series is defined by
$$
\langle \overline{F(x)}F(y)\rangle=\rho(\eta^{-1}(x))+\rho(\eta^{-1}(y))-\rho(\eta^{-1}(x-y)),\qquad x,y\in\mathbb{N},
$$
where $\rho(x)$ is given by (\ref{rho1}) and $\eta$ is the Monna map (\ref{Monna}).

Therefore the correlation function has approximately the power form $|x-y|^{2\alpha-1}$.

Moreover we obtain the deviation from the power behavior related to the stairway structure of the correlation function $\rho(\eta^{-1}(x))$ (since the function $\rho(x)$ is locally constant outside a vicinity of zero).

Stochastic process with the correlation function of the above form can be considered as a model of sandpile avalanche
processes \cite{Avalanche1} and punctuated equilibrium,  similar to applied for the description of biological evolution in \cite{Avalanche}.

Recall that punctuated equilibrium is the property of biological evolution when the species are generally stable and in some
short time periods perform considerable reorganizations, moreover, evolution of one specie triggers the evolution of ecologically related species.
It was proposed to describe punctuated equilibrium by avalanches in a sandpile, where an avalanche triggers other avalanches.

In our model punctuated equilibrium is a result of the structure of time --- the time is $p$-adic with totally disconnected topology,
we consider the time as a real parameter by application of the Monna map $\eta$. After this operation locally constant functions of
$p$-adic parameter become piecewise continuous functions of real parameter. The power $|x-y|^{2\alpha-1}$ in the
correlator becomes a kind of punctuated power function. Our model may be used to support the argument that the biological time
need not be directly represented by real numbers. Thus the biological evolution is in fact continuous and only usage of the physical time makes
the impression of evolutionary discontinuity. We shall elaborate this idea in more detail somewhere else.

As was emphasized the stochastic processes with the correlation function of the above form can be considered as a model of sandpile avalanche
processes. This model can also be applied to modeling of time series obtained as the result of measurement of pressure in oil wells where
a spike in pressure generates a cascade of pressure spikes. The cascade processes have the internal hierarchic and hence multi-fractal structure.
The rings of $p$-adic integers $\mathbb{Z}_p$ are examples of fractals endowed with the algebraic structure and the metric topology. These features
provide the unique possibility to develop analysis and methods for analytic calculations. Thus, although usage of $p$-adics means the lost of
generality in application of the fractal approach, it gives the possibility to find exact analytic answers, as, e.g., the correlation
function which was calculated in this paper.
With a view to wavelet applications for physical and geophysical signals and images decomposition
during their processing, some interesting constructions with linear non-commuting operators $F_l$
(the essential compatibility condition is that the operators form a partition of unity) and,
therefore, {\it non-abelian Cuntz algebra}, was designed by Jorgensen \cite{Jorgensen1}, \cite{Jorgensen2}
in order to select the wavelet packets from libraries of bases. The author consider that these are
constructions which make a selection of a basis with the best frequency concentration in signal or
data compression problem and, which are especially useful for multi-resolution analysis of complex systems \cite{Jorgensen1}, \cite{Jorgensen2}.
From our point of view, this technique is more comparable with our $p$-adic modeling.
A structure theorem was proven by Jorgensen for certain classes of induced scalar measures on the set $X$
which in the case of applications may be the unit interval, or a Cantor set (Dutkay and Jorgensen \cite{Dutkay}),
or an affine fractal etc. Some other researchers have designed the wavelet formalism for multi-fractal analysis
(Murguia and Urias \cite{Murguia}). In our present application of the designed $p$-adic wavelets transform to real data
of seismic exploration and images analyses, see Fig.1, the comparison of our results with the Jorgensen structural model is in view.
We plan to do this in one of our further publications.

\bigskip

\noindent{\bf Acknowledgments}\qquad
This paper was financially supported by the Swedish Scientific Research Council/Swedish Research Link, the project ''Non-Archimedean analysis: from
fundamentals to applications'' and the grant ''Math Modeling of Complex Hierarchical Systems'', Linnaeus University, and
the Projects  N 168638 and  N 143927 of the SENER-CONACYT-Hidrocarburos Research Program.

One of the authors (S.K.) was partially supported by the grant of the Russian Foundation for Basic Research
RFBR 11-01-00828-a, by the grant of the President of Russian
Federation for the support of scientific schools NSh-864.2014.1,  and
by the Program of the Department of Mathematics of the Russian
Academy of Sciences ''Modern problems of theoretical mathematics''.

\end{document}